\documentclass[aps,prl,twocolumn,amsmath,amssymb,floatfix,superscriptaddress]{revtex4}
\usepackage[T1]{fontenc}
\usepackage[latin1]{inputenc}
\usepackage{amsmath}
\usepackage{graphicx}
\usepackage{epsfig}
\usepackage{bm}
\newcommand{\nn}{\nonumber}

\newcommand{\bea}{\begin{eqnarray}}
\newcommand{\ea}{\end{eqnarray}}
\newcommand{\beq}{\begin{equation}}
\newcommand{\eq}{\end{equation}}
\newcommand{\bc}{\begin{center}}
\newcommand{\ec}{\end{center}}
\newcommand{\dg}{\dagger}
\newcommand{\la}{\langle}
\newcommand{\ra}{\rangle}
\begin{document}

\title{Non-equilibrium Josephson oscillations in Bose-Einstein condensates
without dissipation}

\author{Mauricio Trujillo-Martinez}
\affiliation{Physikalisches Institut and Bethe Center for Theoretical Physics,
  Universit\"at Bonn, Nussallee, 12, D-53115 Bonn, 
Germany}

\author{Anna Posazhennikova}
\email{anna.posazhennikova@uni-konstanz.de}

\affiliation{Physikalisches Institut and Bethe Center for Theoretical Physics,
  Universit\"at Bonn, Nussallee, 12, D-53115 Bonn, 
Germany}

\affiliation{Fachbereich Physik, Universit\"at Konstanz, Konstanz, D-78457, Germany}

\author{Johann Kroha}

\affiliation{Physikalisches Institut and Bethe Center for Theoretical Physics,
  Universit\"at Bonn, Nussallee, 12, D-53115 Bonn, 
Germany}


\date{\today}

\begin{abstract}
We perform a detailed quantum dynamical study of non-equilibrium Josephson
oscillations between interacting Bose-Einstein condensates confined in a
finite-size double-well trap. 
We find that the Josephson junction can sustain multiple undamped 
Josephson oscillations up to a characteristic time scale $\tau_c$
without quasipartcles being excited in the system. This may explain 
recent related experiments. Beyond a characteristic time scale 
$\tau_c$ the dynamics of the
junction is governed by fast, quasiparticle-assisted 
Josephson tunneling
as well as Rabi oscillations between the discrete
quasiparticle levels. We predict that an initially self-trapped BEC state will 
be destroyed by these fast dynamics.
\end{abstract}
\maketitle

One of the striking manifestations of quantum mechanics on a macroscopic
level is the particle current induced by the phase difference between
two coherent wave functions connected by a weak link, known as the 
Josephson effect \cite{Josephson62}. For tunneling between the macroscopic
wave functions of two Bose-Einstein condensates (BEC) trapped in a double-well 
potential this phenomenon leads to temporal oscillations of the population 
imbalance $z$ between the two condensates even in the ground state
\cite{Javanainen86,Jack96}. 
However, unlike in superconducting Josephson junctions, 
the interaction between the atoms in the condensates gives rise 
to regimes of fundamentally new dynamical behavior. 
If the initial population imbalance exceeds a critical value
depending on the interaction strength, the large-amplitude Josephson
oscillations (delocalized regime) cease, and the BEC is trapped in one of
the two wells with only small-amplitude oscillations of $z$ about the
non-zero mean value (self-trapped regime). This complex, non-linear 
dynamics has been theoretically predicted for the ground state 
\cite{Milburn97,Smerzi97,Raghavan99} and experimentally
observed recently \cite{Albiez05}. It is not only interes\-ting in 
its own right but also relevant for any merging process of BECs,
e.g., for quantum engineering or for producing 
a continuous source of condensed atoms \cite{Chikkatur02,Yi05}.

In the recent experiment \cite{Albiez05} the Josephson junction is
prepared in a non-equilibrium situation by ramping up the barrier 
between the condensates suddenly, in a non-adiabatic way. 
For this case an immediate damping of the Josephson 
oscillations has been predicted due to quasiparticle excitations
\cite{Zapata03}, which is, however, not observed in the small traps of 
Ref.~\cite{Albiez05}, revealing an incomplete understanding of the
non-equilibrium dynamics.

\begin{figure}[!bt]
\begin{center}
\includegraphics[width=0.32\textwidth]{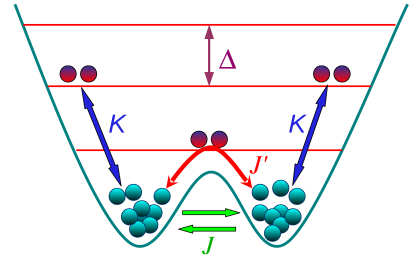}
\end{center}
\vspace*{-0.8em} 
\caption{\label{setup} 
(Color online).
A BEC in a double-well potential after abrupt
decrease of the barrier height. The definitions of the 
parameters are explained in
the text, see Eqs.~(\ref{ham_bec})--(\ref{ham_mix}).
}
\end{figure} 

In this Letter we present a detailed study of the temporal non-equilibrium
dynamics of Josephson-coupled BECs after non-adiabatically 
switching-on the Josephson coupling, including interatomic interactions
as well as quasiparticle (QP) excitations. As the main result we 
find that in small traps multiple, undamped Josephson oscillations are
possibe up to a time scale $\tau_c$. At this time scale the 
dynamics switches abruptly but continuously 
from slow Josephson to fast Rabi oscillations between 
the discrete QP levels. We also predict that the self-trapped
behavior is destroyed by the Rabi oscillations, i.e. for time  $t>\tau_c$
the system switches to delocalized behavior, if it has previously been in a
self-trapped state. This highly non-linear behavior results essentially
from a separation of energy scales in small traps with discrete 
QP level spacing $\Delta$, which can be chosen larger than $J$. 
Switching on $J$ lowers the ground state energy by the amount 
$\Delta E = J\sqrt{N_1(0)N_2(0)}$, where $N_1(0)$, $N_2(0)$ are the 
occupation numbers of the two BECs in the initial state at time $T=0$. 
Thus, after a sudden switching two initially separated BECs are in an 
excited state $\Delta E$ above the coupled ground state. 
Because of the large values of 
$N_1$, $N_2$ this energy is sufficient to excite  QPs
out of the BECs. The time-dependent BEC amplitude acts as a perturbation
on the QP system. However, transitions to QP states 
are not allowed in perturbation theory, because the frequency of the
oscillations is less than their excitation energy, $J<\Delta$.
Our detailed calculations show that such transitions are only 
possible as a highly non-linear process after the characteristic time 
$\tau_c$.

We consider a Bose-Einstein condensed atomic gas in a double-well trap
as represented by Fig. \ref{setup}. Such a system is most generally
described by the Hamiltonian
\bea
H=\int d^3 r \hat \Psi^{\dagger}({\bf r},t)\left(-\frac{1}{2m}\Delta+V_{ext}({\bf
    r},t)\right)\hat \Psi({\bf r},t) \nn \\
+\frac{g}{2}\int d^3 r \hat \Psi^{\dagger}({\bf r},t)\hat \Psi^{\dagger}({\bf
  r},t)\hat \Psi({\bf r},t)\hat \Psi({\bf r},t),
\label{gen_ham}
\ea
where $\Psi({\bf r},t)$ is a bosonic field operator, and we assume a contact interaction between the bosons with $g=4\pi a_s/m$
($a_s$ is the s-wave scattering length). $V_{ext}$ is the 
external double-well trapping
potential. Initially, the barrier between the two wells is
assumed to be infinitely large, so that Josephson tunneling is absent. 
All bosons are condensed, and both condensates are
in the equilibrium state. 
At time $t=0$ the barrier is suddenly lowered so that a Josephson weak link is
established between the wells. This nonadiabatic process drives the system out
of thermodynamic equilibrium.

In order to develop the general non-equilibrium theory for this system
and to analyze its dynamics, we wish to represent the Hamiltonian 
(\ref{gen_ham}) in the complete basis of the exact single-particle eigenstates 
of the double-well potential $V_{ext}({\bf r}, t>0)$ after switching on the
coupling $J$. In this basis the field operator reads, 
\beq
\hat \Psi({\bf r},t)=\phi_1({\bf r})a_1(t)+\phi_2({\bf
  r})a_2(t)+\sum_{n\neq 0}\varphi_n({\bf r}) \hat b_n(t),
\label{wave_function}
\eq
where $\phi_1({\bf r})$, $\phi_2({\bf r})$ are the respective 
ground state wavefunctions of the two wells after lowering the barrier,
and the $a_{\alpha}$ are the corresponding, time-dependent condensate
amplitudes (c-numbers), 
$a_{\alpha}(t)=\sqrt{N_{\alpha}}e^{i\theta_{\alpha}(t)}$, $\alpha=1,2$.
This semiclassical treatment of the BECs neglects phase fluctuations. It
is justified for the experiments \cite{Albiez05}, where
the BECs are initially produced with fixed phase relation and the 
particle number is sufficiently large. The applicability of the
semiclassical approximation has been discussed in detail in 
Refs.~\cite{Zapata03,Zapata98,Pitaevskii01,Xiong06} and has been tested 
experimentally in Ref.~\cite{Esteve08}. The quasiparticle dynamics will
be treated fully quantum mechanically.
The index $n\neq 0$ enumerates the exact single-particle
excited states, with $\varphi_n({\bf r})$ and $\hat b_n(t)$ the 
corresponding eigenfunctions and bosonic destruction operators, respectively.
Note that by including excited states we go beyond the frequently used 
two-mode approximation \cite{Milburn97,Yi05} for the BECs.
For simplicity we assume that the ground state energies of 
the two wells before mixing are equal, $E_0=0$, and that the 
wavefunctions of the excited states extend over both wells. 
Inserting the field operator \eqref{wave_function} into Eq.~(\ref{gen_ham})
and evaluating the overlap matrix elements in a straightforward way, the 
Hamiltonian takes for $t>0$ the form, 
$H=H_{BEC}+H_{qp}+H_{mix}$.
\begin{figure}[!bt]
\begin{center}
\includegraphics[width=0.45\textwidth]{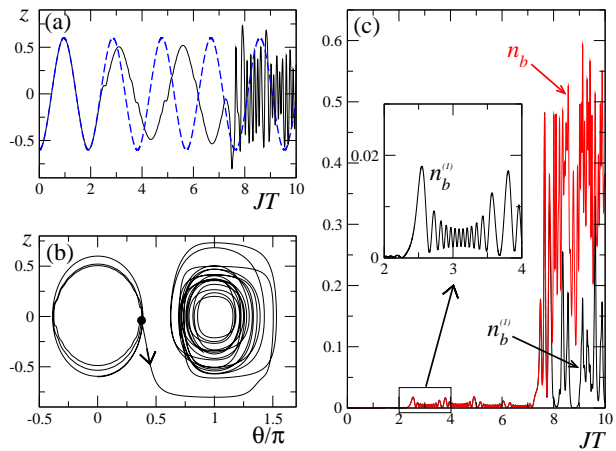}\vspace*{-0.8em} 
\end{center}
\caption{\label{NST} 
(Color online).
Time-evolution of condensed and non-condensed particles for the
initial conditions $z(0)=-0.6$, $\theta(0)=0$ and interaction parameters
$u=u'=5$, $j'=60$, $k=0$ (delocalized regime). 5 QP levels were 
included in the numerical evaluation. 
(a) The dynamics of the BEC population imbalance $z(T)$ is shown (solid black
line). The dashed, blue line shows, for comparison, 
the behavior without QP coupling, $u'=j'=k=0$, 
in agreement with Ref.~\cite{Smerzi97}. 
(b) $z(T)$ vs. $\theta(T)$ map. The arrow indicates the direction of 
time evolution. The time $T=\tau_c$ is marked by the black dot.
It is seen that at $T=\tau_c$  the system 
changes dynamically from a $\theta=0$ to a $\theta=\pi$ junction with
more erratic phase evolution.
(c) Time evolution of the non-condensed particle population.
The black curve (also in the inset) shows the 
particle occupation of the first level $n_b^{(1)}$, while the red curve is the
sum of all five levels $n_b$. 
For $T<\tau_c$ the two curves practically coincide.}   
\end{figure} 
$H_{BEC}$ describes condensate particles,
\beq
H_{BEC}=E_0\sum_{\alpha=1}^2a_\alpha^*a_\alpha+\frac{U}{2}\sum_{\alpha=1}^2
(a_\alpha^{*}a_\alpha^{*}a_\alpha a_\alpha)-J(a_1^*a_2+a_2^*a_1),
\label{ham_bec}
\eq
with the Josephson coupling $J$ and the 
interaction between condensed particles, $U>0$. 
$H_{qp}$ corresponds to single-particle excitations,
\beq
H_{qp}=\sum_{n\neq 0} E_n \hat b_n^{\dg}\hat b_n+\frac{U'}{2}\sum_{n,m}\hat b_m^{\dg} \hat b_n^{\dg}
\hat b_n^{\phantom{\dg}} \hat b_m^{\phantom{\dg}},
\label{ham_qp}
\eq
where $E_n$ are the (bare) QP energies, and  
$U'$ is the repulsive interaction between non-condensed particles.
Mixing between the BECs and the QP system is described by
\bea
H_{mix}&\!\!=\!\!&J'\sum_n \left[(a_1^{*}a_2+a_2^*a_1)\hat b_n^{\dg} \hat
b_n+ \frac{1}{2}(a_1^{*}a_2^{*}\hat b_n \hat b_n+h.c.)\right] \nn \\ 
&\!\!+\!\!&
K\sum_{n,\alpha=1}^{2}\left[\left(a_\alpha^{*}a_\alpha\right)\hat b_n^{\dg}\hat b_n+\frac{1}{4}(a_{\alpha}^{*}a_{\alpha}^{*}\hat
b_n \hat b_n+h.c.)\right]. \nn \\
\label{ham_mix}
\ea
Here the coupling constant $J'$ arises as a
QP-assisted Josephson tunneling as well as a pairwise QP 
creation/destruction out of both BECs {\it simultaneously}. 
$K$ represents the density-density interaction of condensed and non-condensed 
particles and the pairwise QP 
creation/destruction out of each of the BECs {\it separately}. 
In deriving Eq.~\eqref{ham_mix} we
neglected the off-diagonal in $n$ and $m$ elements  because of different
spatial dependence of the wavefunctions. 

To treat the non-equilibrium quantum dynamics of the system, we 
use the Keldysh Green's function ${\bf G}+{\bf C}$, 
generalized to Bose-condensed systems. 
The QP part reads,
\bea
{\bf G}_{nm}(t,t')&=&-i\left(\begin{array}{cc}
\la T_{C}\hat b_n(t)\hat b^{\dg}_m(t')\ra & \la T_{C}\hat b_n(t)\hat b_m(t')\ra \\
\la T_{C}\hat b_n^{\dg}(t)\hat b^{\dg}_m(t')\ra & \la T_{C}\hat b_n^{\dg}(t)\hat b_m(t')\ra
\end{array}
\right) \nn \\
&=&\phantom{-i}\left(\begin{array}{cc}
G_{nm}(t,t') & F_{nm}(t,t') \\
\overline{F}_{nm}(t,t') & \overline{G}_{nm}(t,t'),
\end{array}
\right)\ ,
\label{GF}
\ea
where $T_{C}$ implies time ordering along the Keldysh contour, i.e. each of 
the normal and anomalous bosonic Green's functions $G$ and $F$ 
is a $2\times 2$ matrix in Keldysh space. The condensate part is classical
with trivial time ordering, 
\beq
{\bf C}_{\alpha\beta}(t,t')=-i\left( \begin{array}{cc}
a_\alpha(t)a_\beta^*(t') & a_\alpha(t) a_\beta(t') \\
a_\alpha^*(t)a_\beta^*(t')  & a_\alpha^*(t) a_\beta(t')
\end{array}
\right).
\label{cond}
\eq

The equations of motion for ${\bf G} + {\bf C}$ are derived in a 
standard way \cite{Kadanoff}. 
Transforming the time variables 
to center-of mass and relative coordinates, $T=(t+t')/2$
and $\tau=(t-t')$, observing that the Josephson dynamics 
depending on $T$ is slow compared to the inverse QP 
energies ($J<\Delta$), the relative coordinate can be set $\tau =0$
in all propagators and self-energies. 
We treat the QP interaction in Eq.~\eqref{ham_qp} within the 
self-consistent Bogoliubov-Hartree-Fock approximation \cite{Griffin96}. This 
will be sufficient for the present purpose, since QP collisions,
neglected here, will play a role only for sufficiently high population of
QP levels (see below). The normal and anomalous QP self-energies,
$\Sigma_n (\tau=0,T)$, $\Omega_n (\tau=0,T)$, are then diagonal in the
QP level index $n$ and read,  
\bea
\hspace*{-0.5cm}\Sigma_n (T)&=&K(N_1+N_2)+J'(a_1^*a_2+a_2^*a_1)\nn \\
&+&2iU'\sum_m G_{mm}^{<}(T),  \\
\hspace*{-0.5cm}\Omega_n (T)&=&\frac{K}{2}\sum_{\alpha=1}^2a_\alpha a_\alpha
+J'a_1a_2+iU'\sum_m F_{mm}^{<}(T),\, 
\label{selfenergy}
\ea
with $\overline\Sigma =\Sigma$ and $\overline\Omega =\Omega^{*}$. 
After lengthy but straightforward calculations
one arrives at the coupled set of equations
for the non-condensate propagators $G^{<}(\tau=0,T)$, $F^{<}(\tau=0,T)$  
and the complex condensate amplitudes $a_1(T)$, $a_2(T)$, 
\bea
i\frac{\partial}{\partial T}G^{<}_{nn}(T)&=&\Omega_{n}(T)\overline{F}^{<}_{nn}(T)-\overline\Omega_{n}(T)F^{<}_{nn}(T),\,\,\nn  \\
\Big(i\frac{\partial }{\partial T}-2E_n &-&
2\Sigma_{n}(T)\Big)F_{nn}^{<}(T)=\Omega_{n}(T)\overline{G}^<_{nn}(T)\nn  \\
&+&\Omega_{n}(T)G_{nn}^{<}(T), 
\label{bos_fin}
\ea
\bea
i\frac{\partial}{\partial T}a_1(T)=\left[U|a_1(T)|^2+KN_b(T)\right]a_1(T)-J a_2(T) \nn
\\
+J'N_b(T)a_2(T)+i\left[\frac{K}{2}a_1^*(T)+\frac{J'}{2}a_2^*(T)\right]\sum_n
F^{<}_{nn}(T).
\label{cond_fin}
\ea  
The equation for $a_2(T)$ is obtained from Eq. \eqref{cond_fin} by
$a_1\rightleftarrows a_2$. From Eqs.~\eqref{bos_fin}, \eqref{cond_fin} 
we compute the occupation numbers for bosons out of 
condensate, $N_b^{(n)}(T)=\la \hat b^{\dg}_n(T^+) \hat b_n(T) \ra$, 
$N_b(T)=\sum_n N_b^{(n)}(T)$, the condensate
population imbalance $z(T)=[{N_1(T)-N_2(T)}]/{N}$, 
normalized by the total particle number $N=N_1(0)+N_2(0)+N_b(0)$,
and the time evolution of the phase difference
$\theta(T)=\theta_2(T)-\theta_1(T)$. To absorb the large
factors of particle numbers appearing in 
Eqs.~\eqref{selfenergy}--\eqref{cond_fin} it is useful to define
the dimensionless parameters $u={NU}/{J}$, $u'={NU'}/{J}$, $j'={NJ'}/{J}$, 
$k={NK}/{J}$, and $n_b^{(n)}(T)=N_b^{(n)}(T)/N$.
\begin{figure}[!b]
\begin{center}
\includegraphics[width=0.45\textwidth]{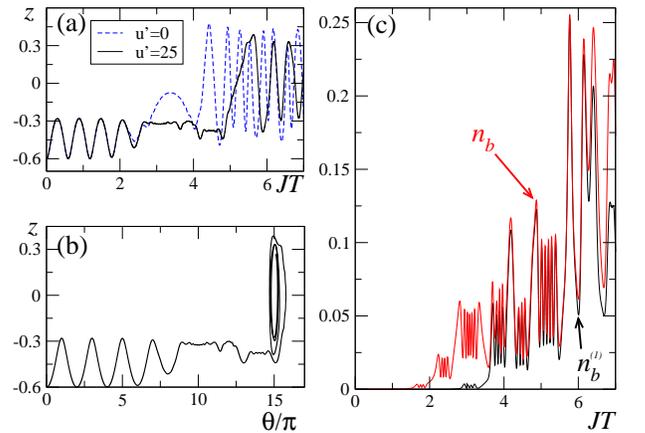}\vspace*{-0.8em} 
\end{center}
\caption{\label{ST}
(Color online).
Same as in Fig. 2, but for $u=u'=25$ (self-trapped regime) and $j'=30$.
In (a) the dashed, blue line shows the behavior without QP
interactions, $u'=0$.}
\end{figure} 

Without coupling to the QP excitations ($j'=k=0$), 
Eq.~\eqref{cond_fin} reduces to the two-mode model of 
Smerzi {et al.} \cite{Smerzi97}, exhibiting the
self-trapped and delocalized regimes, with a Josephson oscillation 
frequency of 
\beq
\omega_J^{(0)} = 2|J| \sqrt{1+u/2}
\label{omega_J}
\eq
in the linear regime (Eq.~(10) in Ref.~\cite{Smerzi97}). 
When, however, $j'\neq 0$, $k\neq 0$ and QPs are excited, $N_b(T)>0$,
the QP-assisted Josephson 
tunneling term in $H_{mix}$ becomes active ($J'$-term in Eq.~\eqref{cond_fin}).
One then expects an enhanced Josephson frequency,
with roughly $J$ replaced by $J[1-j'n_b(T)]$ in Eq.~\eqref{omega_J}. 
At the same time, Rabi oscillations of the $N_b^{(n)}(T)$, i.e. of QP pairs  
between the BEC and the excited levels, with 
frequencies $\omega_R \approx 2E_n$ set in, 
c.f. Eq.~\eqref{bos_fin}. As a result, in this QP-dominated 
regime one expects   
complex, high-frequency anharmonic oscillatory behavior.

The complete numerical solutions of 
Eqs.~\eqref{selfenergy}--\eqref{cond_fin} for a 
finite-size trap with $N=5\cdot 10^5$ particles 
(level spacing $\Delta = E_{n+1}-E_n = 10J$, taking 5 QP levels into 
account \cite{remark}) are shown for typical parameter values
in Fig.~\ref{NST} for the delocalized regime 
and in Fig.~\ref{ST} for the initially self-trapped regime. 
\begin{figure}[!bt]
\begin{center}
\includegraphics[width=0.234\textwidth]{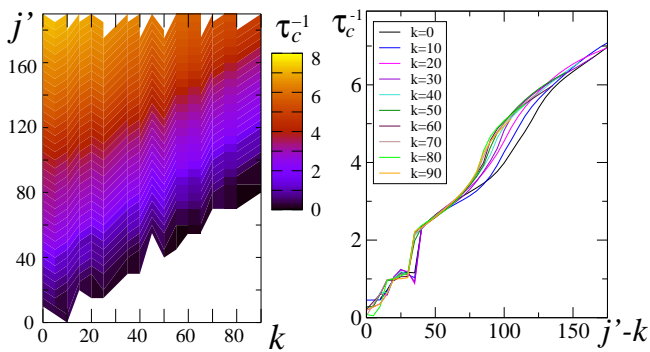}
\includegraphics[width=0.236\textwidth]{figure4_2.eps} 
\vspace*{-0.8em} 
\end{center}
\caption{\label{diag}
(Color online). 
Dependence of the inverse time-scale $\tau_c^{-1}$ on two main parameters $j'$
and $k$, for the initial conditions as in Fig. 3. In the left panel the white
area corresponds to $j'$ and $k$ values for which $\tau_c$ was not
found; we assume $\tau_c \rightarrow \infty$ in this case. 
The right panel shows a collapse of all $\tau_c^{-1}$ curves onto a single 
one with the simple law $\tau_c^{-1}\sim j'-k$. The scatter is due to the
ambiguity in the numerical definition of~$\tau_c$.}
\end{figure} 

The results reproduce the expected behavior discussed above in the
regime with finite QP population.
The parameters were chosen such that the energy $\Delta E$ stored in the 
two BECs by the initial, non-adiabatic switching-on of $J$ is much 
greater than the QP energies, $\Delta E \approx J\sqrt{N_1(0)N_2(0)}
\gg \Delta$. 
The most striking and most important feature seen in both figures is
that nevertheless multiple undamped Josephson oscillations occur 
for an extended period of time without QPs being excited.
The reason for this behavior is that
in the initial state the QP population $n_b(t)$ is vanishing and, therefore,
the QP-assisted Josephson tunneling term $J'$ in Eq.~(\ref{ham_mix}) does
not contribute. Hence, the Josephson oscillations have the bare 
frequency $\omega_J \approx \omega_J^{(0)}<2\Delta$ 
which is not sufficient to excite a 
QP pair perturbatively. Only for times greater than a characteristic 
time $\tau_c$ the highly non-linear dynamics of the system makes QP 
excitations possible. In this long-time regime the finite QP population
$n_b(t)$ and fast oscillations of the BEC population imbalance $z(t)$
stabilize each other mutually: $n_b(t)>0$ implies a QP-enhanced 
Josephson frequency, and the resulting fast oscillations 
($\omega_J>2\Delta$) of $z(t)$ can efficiently excite QPs via 
the mixing Hamiltonian \eqref{ham_mix}. 

The fast, QP-induced dynamics implies two further features. 
(1) As seen from Fig.~\ref{ST} (a) an initially self-trapped state is 
destroyed and the system changes to a delocalized state
at the same time when $n_b(T)$ becomes sizeable. 
(2) At the onset of the fast dynamics the system changes from a 
$\theta=0$ to $\theta=\pi$ Josephson junction, see Figs.~\ref{NST} (b), 
\ref{ST} (b). This can be understood qualitatively, in that the large
phase difference $\theta(T)\approx \pi$ is required to sustain the
large Josephson current in the state with fast dynamics. 

Since the transition to the QP-dominated regime is not described by
a Fermi golden rule, it is hard to analyse the time scale $\tau_c$
analytically. 
We defined $\tau_c$ numerically as the scale where $n_b(t)$ 
first exceeds 0.05 and extracted it from our solutions. 
As seen in Fig.~\ref{ST} (a), $\tau_c$ is essentially independent of the 
QP interaction $u'$. The dependence of $1/\tau_c$ on the 
parameters $j'$ and $k$ (Fig.~\ref{diag}) is remarkably linear, and for
$j'<k$ no transition to the QP-dominated regime is found.

To conclude, we have presented a detailed quantum dynamical 
study of the non-linear Josephson dynamics of BECs confined in a 
finite-size double-well potential,
including coupling to quasiparticle states.
Remarkably, the system can sustain multile, undamped Josephson oscillations
for an extended time period before quasiparticles get excited and the
behavior changes abruptly to a regime of fast Josephson and 
Rabi oscillations.  
Only in this quasiparticle-dominated regime we expect strong 
damping of the oscillations due to inelastic quasiparticle collisions, 
equilibrating the system at a finite temperature. This will be a subject of
further research.
The sharp but continuous transition from the Josephson- to the 
quasiparticle-dominated regime should be experimentally observable.

\acknowledgments
We wish to thank M. Eschrig for helpful discussions.  
This work was supported in part
by DFG through grant No. KR1726/1 (M.T.M., J.K., A.P.), 
SFB 608 (M.T.M., J.K.), and SFB 767 (A.P.).


\begin{thebibliography} {99}
\bibitem{remark} 
We have performed similar calculations for 2, 3 and 4 QP levels with 
fixed $\Delta$. It was verified that the generic behavior described in the text,
in particular the values of $\tau_c$, are essentially independent of 
the number of QP levels taken into account.
\end{thebibliography}
\end{document}